\documentclass[manuscript]{emulateapj}

\newcommand{\mum}{${\rm \mu m}$}
 \newcommand{\ergs}{${\rm ergs~s^{-1}}$}
 \newcommand{\msun}{M$_\odot$}
 \newcommand{\lsun}{L$_\odot$}

 \newcommand{\lx}{$L_{\rm X}$}
 \newcommand{\lbol}{$L_{\rm Bol}$}
 \newcommand{\mbulge}{$M_{\rm Bulge}$}
 \newcommand{\mgal}{$M_{\ast}$}
 \newcommand{\mbh}{$M_{\rm BH}$}
 \newcommand{\mdot}{$\dot{M}_{\rm BH}$}
 \newcommand{\smdot}{${\rm s}\dot{M}_{\rm BH}$}
 \newcommand{\redd}{$\lambda_{\epsilon}$}

\def\simgt{\lower.5ex\hbox{\gtsima}}
\def\simlt{\lower.5ex\hbox{\ltsima}}

\shorttitle{The AGN main sequence}
\shortauthors{Mullaney et al.}
\begin{document}

\title{The hidden ``AGN main sequence'': Evidence for a universal
  black hole accretion to star formation rate ratio since
  $\lowercase{z}\sim2$ producing a \mbh--\mgal\ relation}

\author{J. R. Mullaney\altaffilmark{1,2}, E. Daddi\altaffilmark{1},
  M. B\'{e}thermin\altaffilmark{1}, D. Elbaz\altaffilmark{1},
  S. Juneau\altaffilmark{1}, M. Pannella\altaffilmark{1},
  M. T. Sargent\altaffilmark{1}, D. M. Alexander\altaffilmark{2},
  R. C. Hickox\altaffilmark{3}} \altaffiltext{1}{Irfu/Service
  d'Astrophysique, CEA-Saclay, Orme des Merisiers, 91191,
  Gif-sur-Yvette Cedex, France} \altaffiltext{2}{Department of
  Physics, Durham University, South Road, Durham DH1 3LE, U.K.}
\altaffiltext{3}{Department of Physics and Astronomy, Dartmouth
  College, 6127 Wilder Laboratory, Hanover, NH 03755, USA}

\begin{abstract} 

  Using X-ray stacking analyses we estimate the average amounts of
  supermassive black hole (SMBH) growth taking place in star-forming
  galaxies (SFGs) at $z\sim1$ and $z\sim2$ as a function of galaxy
  stellar mass (\mgal).  We find the average SMBH growth rate follows
  remarkably similar trends with \mgal\ and redshift as the average
  star-formation rates (SFRs) of their host galaxies (i.e.,
  \mdot$\propto$\mgal$^{0.86\pm0.39}$ for the $z\sim1$ sample and
  \mdot$\propto$\mgal$^{1.05\pm0.36}$ for the $z\sim2$ sample). It
  follows that the ratio of SMBH growth rate to SFR is (a) flat with
  respect to \mgal\ (b) not evolving with redshift and (c) close to
  the ratio required to maintain/establish a SMBH to \mgal\ ratio of
  $\approx10^{-3}$ as also inferred from today's \mbh--\mbulge\
  relationship.  We interpret this as evidence that SMBHs have, on
  average, grown in-step with their host galaxies since at least
  $z\sim2$, irrespective of host galaxy mass and AGN triggering
  mechanism.  As such, we suggest that the same secular processes that
  drive the bulk of star formation are also responsible for the
  majority of SMBH growth.  From this, we speculate that it is the
  availability of gas reservoirs that regulate both cosmological SMBH
  growth and star formation.

\end{abstract}

\keywords{galaxies: active---galaxies: evolution---galaxies: star
  formation---X-rays: general}

\section{Introduction}
\label{Introduction}

The tight observed relationship between galaxy bulge mass and the mass
of its central, supermassive black hole (SMBH; e.g.,
\citealt{Haring04}), hereafter \mbh--\mbulge, suggests galaxy growth
(i.e., star-formation) is closely tied to the principal mode of SMBH
growth: accretion during periods of nuclear activity (i.e., active
galactic nuclei, or AGN).  However, the \mbh--\mbulge\ relationship
only provides a snapshot of the end result, with the details
surrounding when, how and under what conditions these links were
forged remaining poorly understood.  Indeed, it is not yet clear
whether {\it all} episodes of star-formation are eventually
accompanied by SMBH growth, or whether such evolutionary links are
limited to the most rapidly growing systems, such as those induced by
major-mergers (see \citealt{Alexander11} for a review).  A major
difficulty in exploring the links between ongoing SMBH and galaxy
growth stems partly from scatter introduced by the different duty
cycles of AGN and star-formation episodes, leading to what appears to
be only very weak correlations between the two events (e.g.,
\citealt{Silverman09,Mullaney12}).

Since SMBH growth appears to be so closely tied to galaxy growth, it
is pertinent to ask whether average SMBH accretion rates trace
star-formation rates (SFRs). In this vein, \cite{Daddi07b} showed that
the ratio of average SMBH accretion rate to SFR in star-forming
galaxies (SFGs) at $z\sim2$ was roughly consistent with that inferred
from today's \mbh--\mbulge\ relationship.  However, focussing on only
the global average conceals details of how SMBH and galaxy mass is
built up. Indeed, it is now evident that SFGs have formed stars at a
rate that is roughly proportional to their stellar masses (\mgal)
since at least $z\sim2$, while their average specific SFRs (i.e.,
${\rm sSFR=SFR}$/\mgal) increase strongly with redshift (e.g.,
\citealt{Noeske07,Elbaz07,Daddi07,Pannella09,Karim11,Elbaz11}).  Here,
we determine whether these trends between SFR, \mgal\ and redshift for
SFGs also extend to the growth of their resident SMBHs.  We use
$H_0=71~{\rm km~s^{-1}~Mpc^{-1}}$, $\Omega_{\Lambda}=0.73$,
$\Omega_{\rm M}=0.27$ and a Chabrier initial mass function.

\section{Data and analyses}
\label{Data}

\begin{table*}
\begin{center}
  \caption{Derived Average Physical Properties of Sub-samples}\label{stats}
\begin{tabular}{@{}lcccccccc@{}}
\hline
\hline
(1)&(2)&(3)&(4)&(5)&(6)&(7)&(8)&(9)\\
Mass range&$N_{\rm Det}$&$N_{\rm Stk}$&$\langle{M}_{\ast}\rangle$&$\langle{\rm SFR}\rangle$&$\langle{F_{\rm 2-10keV}}\rangle$&$\langle{L_{\rm 2-10keV}}\rangle$&$\langle{L_{\rm Bol}}\rangle$&$\langle{\dot{M}_{\rm BH}}\rangle$\\
\hline
$z\sim 1$&&&&&&&&\\
$9.76-10.09$ &     12&    138&$9.952\pm0.088$&$3.89\pm0.32$&$20^{+38}_{-16}$&$0.5^{+1.0}_{-0.3}$&$3.3^{+5.8}_{-2.1}$&$2.0^{+3.6}_{-1.3}$\\
$10.09-10.42$&     22&    112&$10.232\pm0.092$&$7.19\pm0.51$&$22^{+15}_{-10}$&$0.71^{+0.94}_{-0.37}$&$4.4^{+5.3}_{-2.4}$&$2.7^{+3.2}_{-1.5}$\\
$10.42-10.75$&     28&     82&$10.592\pm0.098$&$12.62\pm0.97$&$82^{+43}_{-31}$&$3.8^{+2.9}_{-1.6}$&$22^{+17}_{-10}$&$14^{+10}_{-6}$\\
$10.75-11.25$&     41&     77&$10.95\pm0.14$&$16.0\pm1.6$&$82^{+48}_{-33}$&$3.7^{+2.5}_{-1.4}$&$22^{+14}_{-8}$&$13.3^{+8.5}_{-5.2}$\\
$z\sim 2$&&&&&&&&\\
$9.76-10.09$ &     20&    327&$9.907\pm0.094$&$20.77\pm0.89$&$9.9^{+7.6}_{-5.1}$&$2.8^{+2.8}_{-1.3}$&$17^{+16}_{-8}$&$10.3^{+9.6}_{-5.0}$\\
$10.09-10.42$&     32&    206&$10.24\pm0.10$&$34.1\pm2.0$&$17.3^{+7.2}_{-5.6}$&$4.0^{+2.9}_{-1.5}$&$24^{+16}_{-10}$&$14.7^{+9.7}_{-5.8}$\\
$10.42-10.75$&     34&     67&$10.564\pm0.091$&$58.4\pm5.3$&$50^{+24}_{-18}$&$12.9^{+8.8}_{-4.9}$&$77^{+49}_{-30}$&$47^{+30}_{-18}$\\
$10.75-11.25$&     18&     28&$10.90\pm0.11$&$151\pm22$&$140^{+130}_{-80}$&$25^{+25}_{-12}$&$150^{+140}_{-70}$&$90^{+86}_{-44}$\\
\hline
\end{tabular}
\end{center}
\tablecomments{(1) Stellar mass range (log$[$\msun$]$), (2) Number of X-ray detected galaxies, (3)
  Number of stacked X-ray undetected galaxies, (4) Stellar
  mass (log$[$\msun$]$),
  (5) SFR (\msun~yr$^{-1}$), (6) Observed-frame
  2-10~keV X-ray flux (${\rm 10^{-17}~ergs~s^{-1}~cm^{-2}}$), (7)
  Intrinsic AGN rest-frame 2-10~keV X-ray luminosity (${\rm
    10^{42}~ergs~s^{-1}~cm^{-2}}$)
  (8) Bolometric AGN luminosity ($10^9$~\lsun), (9) SMBH
  accretion rate ($10^{-3}$~\msun~yr$^{-1}$). Columns 4-9 contain
  mean-average values.}
\end{table*}

We measure the average SMBH accretion rates in SFGs at $0.5<z<2.5$ in
the GOODS-South field.  Our two samples of $z\sim1$ and $z\sim2$
galaxies (607 and 1146 sources, respectively) are from the K-selected
catalogue of \citeauthor{Daddi07} (\citeyear{Daddi07,Daddi07b}; see
also \citealt{Salmi12} for details of the $z\sim1$ sample).  SFRs for
these galaxies are based on 24~\mum\ and UV observations,
respectively, and are known to be unbiased on average
(\citealt{Daddi07,Elbaz10}).  Both samples were divided into the same
set of stellar mass (\mgal) bins.  The average SFR of SFGs in these
bins, plotted as a function of \mgal, is shown in Fig.  \ref{Lx_M}a.
The shallower slope of the $z\sim1$ SFR--\mgal\ relation compared to
the $z\sim2$ sample is due to an Eddington bias introduced by the flux
limit of the 24~\mum\ data used to estimate their SFRs.  By comparing
like-for-like average X-ray emission (and inferred SMBH accretion
rates) with average SFRs and using the SFGs as priors for our X-ray
matching/stacking we ensure that this bias has no effect on our
results.

The X-ray data used for this study were taken from the 4\ Ms {\it
  Chandra} deep-field observations (Cycle 9 DDT; see \citealt{Xue11}
for details), which entirely cover our SFG samples. To determine the
average level of SMBH accretion taking place in the SFGs we account
for X-ray non-detections as well as X-ray detections.  First, we used
positional matching to identify those galaxies detected in X-rays,
matching to the optical positions reported in \cite{Xue11} and
assuming a matching radius of 1\arcsec.  The numbers of identified
matches in each of our \mgal\ and redshift bins are given in Table
\ref{stats}.  For the remainder, we stacked the X-ray data at the
optical positions of the SFGs, taking care to avoid detected sources
and only stacking within 8\arcmin\ of the average aim-point of the
{\it Chandra} observations.\footnote{We note the results from our
  X-ray stacks are consistent within the errors of those obtained
  using CSTACK \mbox{(http://cstack.ucsd.edu/)} developed by Takamitsu
  Miyaji, which uses the 2~Ms CDF-S data.} For each of our redshift
and mass bins the total (i.e., detected $+$ undetected) X-ray counts
are dominated by the X-ray detected sources.  Average count rates were
determined by summing the counts from the detected sources and the
stacks then dividing this by the total effective exposure times (of
both detected and undetected sources).  Average band ratios, fluxes at
the observed-frame 2-10~keV band and obscuration-corrected
luminosities at a rest-frame 2-10~keV band (i.e., \lx) were calculated
using the methodology outlined in \cite{Luo08} which uses band-ratios
to correct for obscuration (see our Table \ref{stats}).  The average
contribution to \lx\ from star-formation was calculated using two
different SFR-\lx\ relations (from \citealt{Ranalli03} and
\citealt{Vattakunnel11}) and subtracted to leave the intrinsic \lx\ of
the AGN. Both relations estimate a non-AGN contribution of $<5\%$ in
each of our mass and redshift bins, meaning this correction has no
significant impact on our results.

Once the average intrinsic X-ray luminosities had been estimated for
the SFGs in each of our \mgal\ and redshift bins, we used this
information to estimate average AGN bolometric luminosities (i.e.,
\lbol).  For simplicity, we derive our main results using a constant
bolometric correction factor of 22.4 to convert \lx\ to \lbol\ (the
median bolometric correction factor of a sample of local,
\lx$=10^{41-46}$~\ergs\ AGN from \citealt{Vasudevan07}).  From \lbol\
we derive SMBH accretion rates (i.e., \mdot) using:

\begin{equation}
\label{mdot}
\dot{M}_{\rm BH}(M_{\ast}, z) = \frac{(1-\epsilon) L_{\rm
    bol}(M_{\ast}, z)}{\epsilon c^2}
\end{equation}
\\
\noindent
where $c$ is the speed of light in a vacuum and $\epsilon$ is the
efficiency by which mass is converted into radiated energy via the
accretion process.  Here we assume $\epsilon=0.1$ (e.g.,
\citealt{Marconi04}), or that roughly 10\% of mass within the
accreting system is converted into energy that is radiated away via
electromagnetic radiation, irrespective of \mbh.

As the number of X-ray counts for each bin is dominated (i.e.,
$>80\%$) by X-ray detected sources, uncertainties on the mean \lx\
were calculated using a bootstrapping technique; repeatedly selecting
2/3 of the detected sample in each bin at random and calculating the
dispersion of the resulting \lx\ distribution.  The uncertainties on
\lbol\ and \mdot\ were then propagated from our estimates of the
uncertainties on the mean \lx.

\section{Results}
\label{Results}

\begin{figure}
\plotone{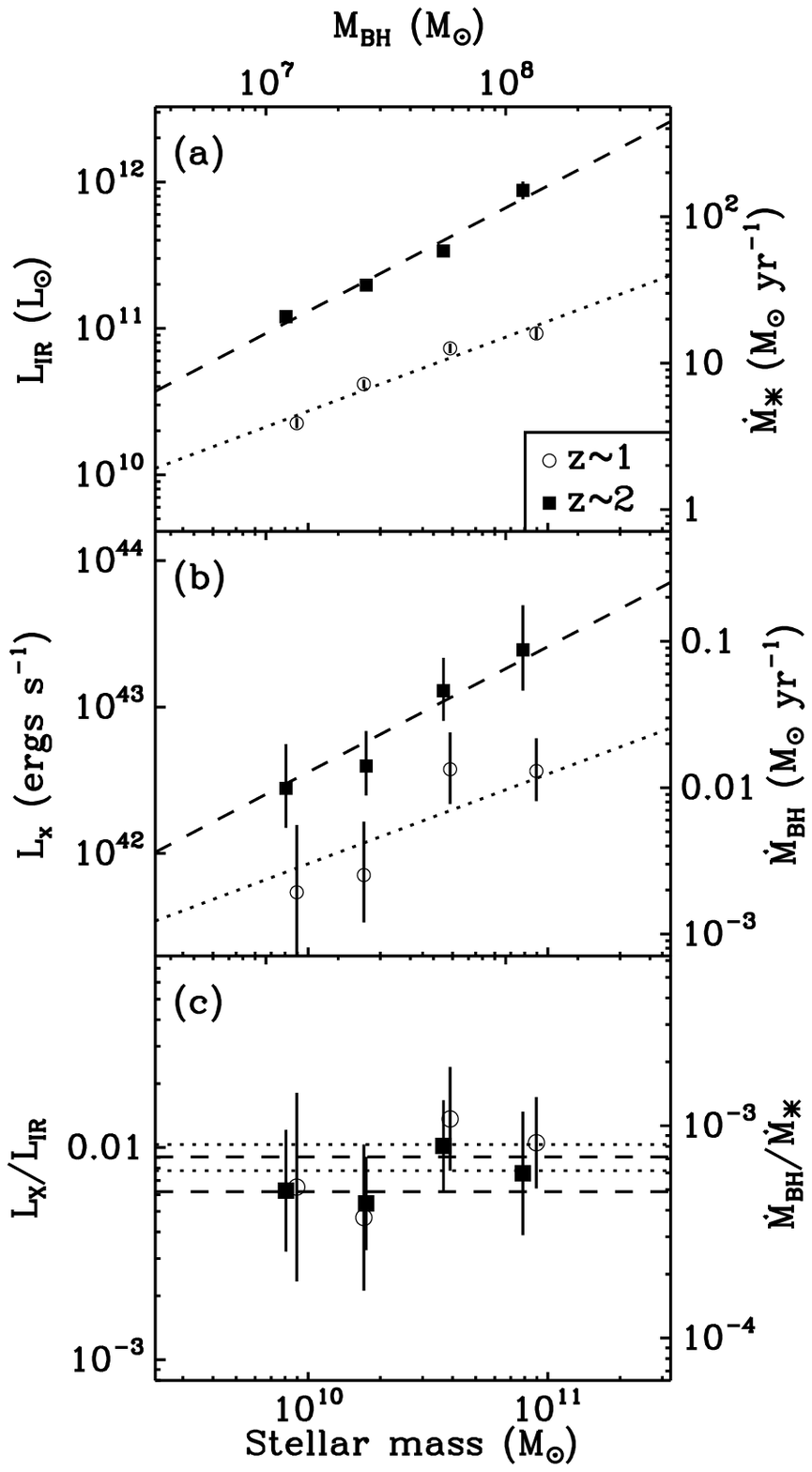}
\caption{\textbf{(a)} Average SFRs (right-hand axis) versus stellar
  mass for our $z\sim1$ (open circles) and $z\sim2$ (filled squares)
  samples of SFGs (left-hand axis gives equivalent infrared luminosity
  for illustrative purposes only).  Dotted and dashed lines indicate a
  least-squares linear fit to these data.  \textbf{(b)} Average X-ray
  luminosities of the SFGs in our samples (same symbols as top panel)
  after accounting for any host galaxy contribution.  Lines have the
  same gradients as in the top panel, only normalised to best-fit the
  inferred \mdot, which is indicated in the right-hand
  axis. \textbf{(c)} Average SMBH accretion rate to SFR ratio for our
  two redshift samples. The uncertainties on these points are
  consistent with a flat \mdot/SFR ratio with respect to \mgal\ for
  both the $z\sim1$ and $z\sim2$ samples, indicated by the dotted and
  dashed lines, respectively. 1-$\sigma$ uncertainties are included in
  each panel, but are smaller than the points in panel (a).}
\label{Lx_M}
\end{figure}

In Fig. \ref{Lx_M}b we plot the average \lx\ of X-ray detected $+$
undetected (i.e., stacked) SFGs as a function of \mgal\ for our two
redshift samples.  Both our $z\sim1$ and $z\sim2$ samples show a clear
increase in their average \lx\ with increasing \mgal\ for the mass
range considered.  This is in contrast to studies of individually
detected X-ray AGN which find no such correlation (e.g.,
\citealt{Mullaney12}).  It is only when the scatter in \lx\ due to AGN
variability is averaged-out that the correlation between \lx\ and
\mgal\ for SFGs presents itself.  A least-squares fit to these data
gives \lx$\propto$\mgal$^{0.86\pm0.39}$ for the $z\sim1$ sample and
\lx$\propto$\mgal$^{1.05\pm0.36}$ for the $z\sim2$ sample.  As we
neglect the (unknown) possible variations of accretion efficiencies
and bolometric corrections with stellar masses, \lx\ can be directly
replaced by \mdot\ in these equations to give the same relationships
between \mdot\ and \mgal.  Importantly, we also find that the average
\lx\ of SFGs increases with redshift, being a factor of $5.2\pm1.4$
higher, on average, at $z\sim2$ compared to $z\sim1$.  This is
comparable to the factor of $6.1\pm2.3$ higher average SFRs of the
$z\sim2$ sample.

To demonstrate this last point we have included in our \lx-\mgal\ plot
(Fig. \ref{Lx_M}b) the observed trend between SFR and \mgal\ derived
from our two samples of SFGs (i.e., SFR$\propto$\mgal$^{0.6}$ and
SFR$\propto$\mgal$^{0.9}$ for the $z\sim1$ and $z\sim2$ populations,
respectively), normalised to fit the average inferred \mdot\ of the
respective redshift sample but maintaining the gradient. Plotting
\mdot/SFR as a function of \mgal\ (Fig. \ref{Lx_M}c) we find this
ratio is only marginally dependent on \mgal\ and is strikingly similar
for our $z\sim1$ and $z\sim2$ samples (i.e.,
\mdot/SFR~$\propto$~\mgal$^{0.3\pm0.4}$ and
\mdot/SFR~$\propto$~\mgal$^{0.2\pm0.4}$ for the $z\sim1$ and $z\sim2$
samples, respectively). Furthermore, the uncertainties are consistent
with a flat \mdot/SFR ratio with respect to \mgal\ for both samples
(i.e., \mdot~$=~[0.6-0.8]\times10^{-3}\cdot$SFR for our $z\sim1$
sample and \mdot~$=~[0.5-0.7]\times10^{-3}\cdot$SFR for our $z\sim2$
sample).\footnote{\label{lumbol} Adopting a luminosity-dependent
  bolometric correction factor from \cite{Hopkins07} gives
  \mdot/SFR~$\propto$~\mgal$^{0.5\pm0.5}$ for both redshift bins; the
  error-bars remain consistent with a flat distribution.}

By taking the average X-ray output of SFGs it follows that the
ensemble growth rate of SMBHs increases with both increasing \mgal\
and redshift in a manner that is remarkably similar to the average
levels of star-formation taking place in SFGs.  The independence of
the average \mdot/SFR ratio on \mgal\ implies its constancy during the
rapid growth phases of galaxies. Next, we consider how this constant
ratio conforms to our understanding of relative SMBH growth both
locally and at high redshifts.

The mass of a SMBH today, at redshift $z_f=0$, can be described in
terms of its total accretion history since $z_i$ and its mass at
$z_i$, i.e.,\footnote{We neglect merging SMBHs as they will not affect
  the {\it total} mass contained within SMBHs while merger-induced
  starbursts contribute only $\sim10\%$ of stellar-mass build-up
  (\citealt{Rodighiero11}).}

\begin{equation}
\label{eq1}
M_{\rm BH}(z_f) = M_{\rm BH}(z_i) + \int^{z=z_f}_{z=z_i}\dot{M}_{\rm BH}(t)dt
\end{equation}
\noindent
Our observations support a constant average ratio between the SMBH and
galaxy growth rates, i.e., $\dot{M}_{\rm BH} = \alpha
\dot{M}_{\ast}$.\footnote{Note: $\dot{M}_{\ast}\equiv{\rm SFR}$}
Replacing the resulting integral with $\Delta
M_{\ast}(z={z_i}\rightarrow{z_f})$, we obtain,
\begin{equation}
\label{eq2}
M_{\rm BH}(z_f) = M_{\rm BH}(z_i) + \alpha\Delta M_{\ast}(z={z_i}\rightarrow{z_f})
\end{equation}
\noindent
Similarly, the stellar mass of the host galaxy at $z_f$ is given by,
\begin{equation}
\label{eq3}
M_{\rm \ast}(z_f) = M_{\rm \ast}(z_i) + \Delta M_{\ast}(z={z_i}\rightarrow{z_f})
\end{equation}
\noindent
so the black hole to stellar mass ratio is given by, 
\begin{equation}
\label{eq4}
\frac{M_{\rm BH}(z_f)}{M_{\ast}(z_f)} = 
\frac{M_{\rm BH}(z_i) + {\alpha}{\Delta}M_{\ast}(z={z_i}\rightarrow{z_f})}{M_{\ast}(z_i) + {\Delta}M_{\ast}(z={z_i}\rightarrow{z_f})}
\end{equation}
\noindent
Defining $\beta$ as the initial \mbh\ to \mgal\ ratio (relative to the
growth rate ratio, i.e., $\alpha$) and $\gamma$ as the relative change
in \mgal, i.e.,
\begin{equation}
\label{eq5}
M_{\rm BH}(z_i) = \beta\alpha{M_{\ast}(z_i)}
\mbox{, }
\gamma = \frac{{\Delta}M_{\ast}(z={z_i}\rightarrow{z_f})}{M_{\ast}(z_i)}
\end{equation}
we obtain, 
\begin{equation}
\label{eq6}
\frac{M_{\rm BH}(z_f)}{M_{\ast}(z_f)} =
\alpha\frac{\gamma+\beta}{\gamma+1}\approx\alpha\ \left({\rm when\
  }\gamma\gg\beta\right)
\end{equation}
Thus, as soon as enough activity has taken place so that the
(uncertain) initial conditions can be neglected, one expects constant
\mbh\ to \mgal\ ratios independent of redshift and roughly equal to
the observed growth rate ratio. It is not surprising then that the growth
ratios are close to the SMBH to stellar mass ratio inferred from
today's \mbh--\mbulge\ relationship, indicating that this relative
growth rate is crucial in defining these ratios.

\section{Discussion: A hidden AGN ``main sequence''}
\label{Discussion}

\begin{figure}
\plotone{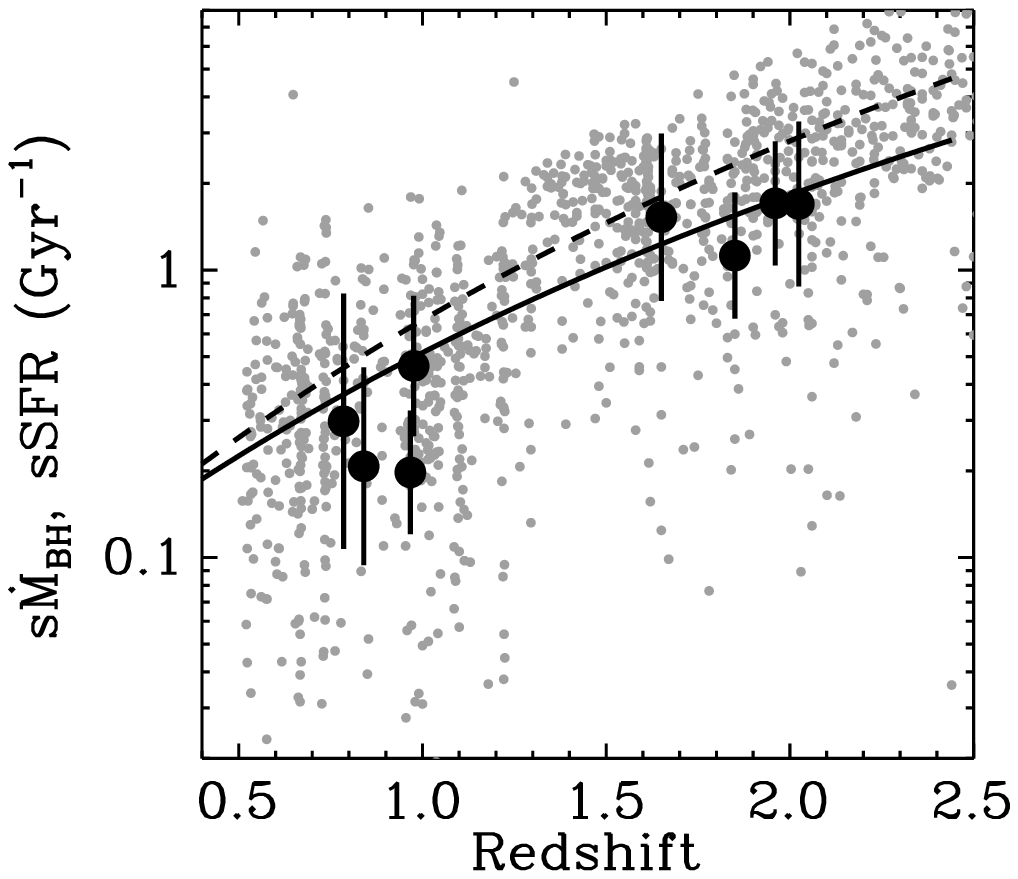}
\caption{Specific SMBH accretion rates (i.e., \smdot$=$\mdot/\mbh;
  assuming $M_{\rm BH}=1.5\times10^{-3}M_{\ast}$) plotted as a
  function of redshift for our stellar mass and redshift bins (large
  black points).  Included in this plot are the sSFRs of the galaxies
  in our samples (small gray points), and the sSFR-$z$
  relationships from \cite{Elbaz11} and \cite{Pannella09} (solid and
  dashed lines, respectively).  We have increased the specific SMBH
  accretion rates by a factor of 2 to account for missing AGN due to
  e.g. obscuration, but note that the {\it relative} change in average
  \mdot/\mbh\ between our redshift bins is remarkably similar to that
  of the sSFRs.}
\label{sMdot}
\end{figure}

Our results suggest that it is coeval growth at constant relative
rates averaged over cosmological timescales that produces the links
between SMBH and stellar mass inferred from the \mbh--\mbulge\
relation.  To address this in more detail there are a number of points
that should be considered carefully.

First, we emphasise that our results are cosmologically relevant,
referring to the bulk of the SMBH and galaxy growth.  The $0.5<z<2.5$
epoch spanned by our samples correspond to the vast majority of both
global star (e.g., \citealt{Dickinson03,Magnelli11}) and SMBH (e.g.,
\citealt{Marconi04}) formation history.

Likewise, although we miss the most luminous AGN their absence will
not change our results.  Integrating the ``LADE'' AGN X-ray luminosity
function of \citealt{Aird10}, we estimate that $\sim$20--30\%\ of all
SMBH accretion at $0.5<z<2.5$ takes place in AGNs that are rare enough
such that $\leq 3$ would be expected to be found in our survey (i.e.,
rarer than three per $2\times10^{5}~{\rm Mpc^3}$ at $z\sim1$ and three
per $4\times10^{5}~{\rm Mpc^3}$ at $z\sim2$, corresponding to
\lx$>2\times10^{44}$~\ergs and \lx$>3\times10^{44}$~\ergs,
respectively).  Similarly, we could miss $\lesssim2\%$ of the SFR
density because of volume effects. We note that the fraction of AGNs
in low-SFR galaxies not included in our SFG sample is also negligible
at these redshifts, being $\lesssim10$\% (e.g., \citealt{Mullaney12}).

Obscuration is a potentially more serious issue, as we will
underestimate the contribution of the heavily obscured (i.e.,
Compton-thick) AGNs thought to be responsible for $\lesssim$50\%\ of
total SMBH growth (e.g., \citealt{Gilli07}). This could introduce a
factor of $\lesssim 2$ correction, but is unlikely to be substantially
larger than the observed, unobscured contribution.  Obscuration due to
orientation effects (unified model) is unlikely to depend strongly on
either mass or redshift and, as such, will not affect the observed
correlations.  It is unclear whether the levels of obscuration due to
merger driven starbursts changes as a function of galaxy mass and/or
redshift.  However, the fraction of starbursts does not appear to
change significantly with redshift or galaxy mass and accounts for
only 10--15\% of all star-formation (at least for the ranges
considered here; \citealt{Rodighiero11,Sargent12}).

This obscured AGN fraction, together with the fact that a fraction of
the stars forming will quickly die due to stellar evolution, leads us
to conclude that our results support a constant \mbh\ to \mgal\ ratio
of:

\begin{equation}
  \frac{{M}_{\rm BH}}{M_{\ast}} \approx (1-2)\times10^{-3} 
\label{Mag}
\end{equation}

\noindent 
at $0.5<z<2.5$ -- consistent with the conclusions of \cite{Jahnke09}
and \cite{Cisternas11b}.  This ratio is also consistent with the local
\mbh/\mbulge\ ratio, suggesting that it is the same relation.  At this
point it is important to note that, while there is some evidence to
suggest that today’s \mbh\ correlates most tightly with bulge mass
(\citealt{Kormendy11}), for the sake of this letter we do not
distinguish between galaxy and bulge mass/SFR as it is impossible to
reliably determine which of the stars formed at $z\gtrsim0.5$ will be
in bulges by $z\sim0$.  Having said that, it is thought the the
majority of stars formed at these high redshifts in the \mgal\ range
considered will collapse to form massive bulges by $z\sim0$ (e.g.,
\citealt{Renzini06}), probably due to the effects of mergers.

\begin{figure}
\plotone{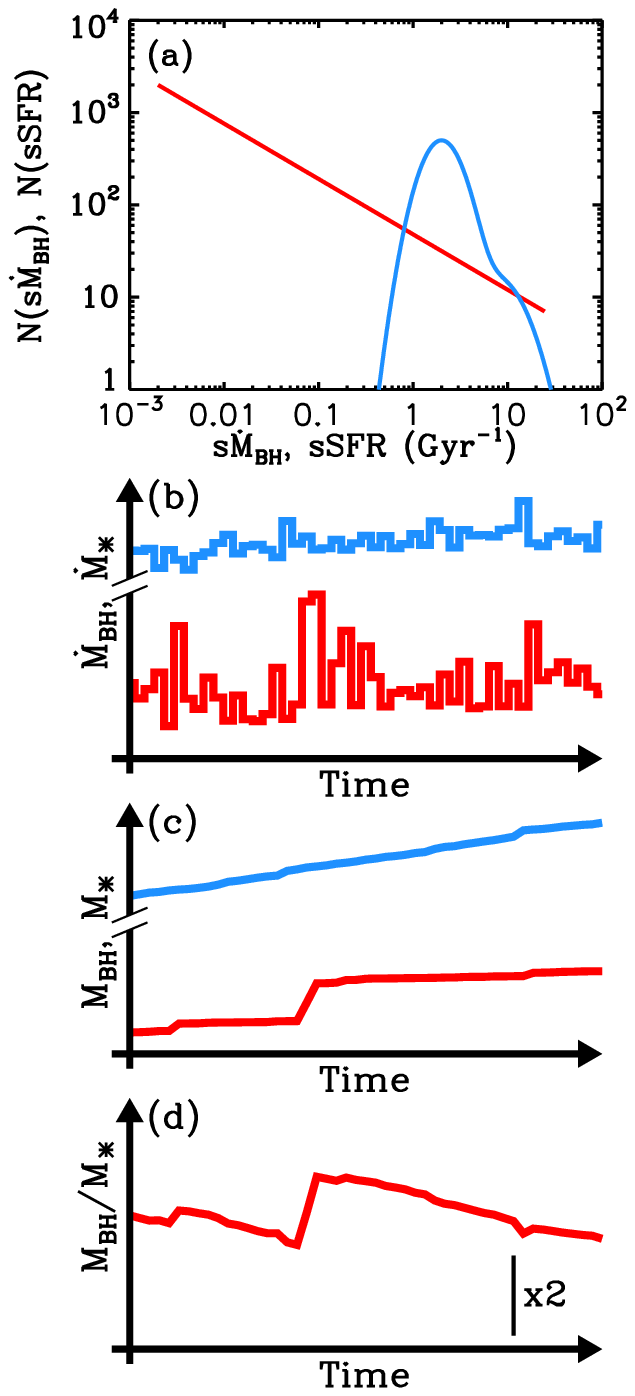}
\caption{\textbf{(a)}: Probability distribution functions (PDFs) of
  \smdot\ ($\propto$ Eddington ratio; red: \citealt{Aird12}) and sSFRs
  (blue; \citealt{Sargent12}) at a given redshift (arbitrary
  $y$-scaling).  Note the broad \smdot\ PDF, indicative of the large
  variations in nuclear activity compared to the sSFR of the
  host. {\it Lower three panels}: Cartoon illustrating the growth
  rates \textbf{(b)} and total and relative masses (\textbf{c} and
  \textbf{d}, respectively) of SMBHs and their hosts.  The host grows
  steadily, whereas the SMBH grows in fits and spurts, causing the
  SMBH mass to ``lead and lag'' the galaxy mass at different times
  but, on average, remaining closely tied.  Scale in panel (d)
  indicates a factor of 2 change.}
\label{GrowthSketch}
\end{figure}

Using Eqn. (\ref{Mag}) we can compute approximate SMBH masses for our
galaxy samples.  Since the \mdot\ to SFR ratio has remained consistent
with the SMBH and galaxy mass ratio since $z\sim2$ the specific SMBH
growth rate (i.e., \smdot$=\dot{M}_{\rm BH}/M_{\rm BH}$) traces the
same trend with redshift as the average sSFRs of SFGs (e.g.,
\citealt{Pannella09, Elbaz11}; Fig. \ref{sMdot}).  Thus, when doing
ensemble (i.e., time) averages, the SMBH population forms an ``AGN
main-sequence'' (where roughly \mdot$\propto$\mbh, on average) that
follows the same trends with stellar mass and redshift as the
so-called galactic main sequence of e.g., \cite{Noeske07, Elbaz07,
  Daddi07}.

It is interesting to interpret these results in terms of the frequency
of nuclear and star-forming activity in galaxies.  For this, we
consider the distribution of AGN Eddington ratios
(\redd$\propto$\mdot/\mbh) and galaxy sSFRs.  Recently, \cite{Aird12}
suggested that the \redd\ distribution of X-ray AGNs can be described
purely as a function of \redd\ and redshift; i.e., independently of
\mgal\ (Fig. \ref{GrowthSketch}a).\footnote{Studies of
  optically-selected, broad-line quasars have reported log-normal
  \redd\ distributions.  However, by selection, those AGNs have
  considerably higher average \lbol\ ($\approx10^{13}$~\lsun; e.g.,
  \citealt{Shen08}) than our samples and, as such, are less directly
  relevant to our analyses.}  This broad distribution for AGNs, which
spans over four orders of magnitude in \redd\ (see also, e.g.,
\citealt{Babic07, Hopkins09}), contrasts with the distribution of sSFR
of galaxies that is remarkably narrow, yet also independent of \mgal\
(\citealt{Sargent12}; Fig. \ref{GrowthSketch}a).  This is the main
reason why the AGN main sequence remains hidden; there are strong
changes in \smdot\ compared to minor changes in the sSFRs of galaxies
(Fig. \ref{GrowthSketch}). This has the implication that outliers
should exist in the $M_{\rm BH}$-\mgal\ relation when the SMBH growth
has taken advantage over the \mgal\ growth and vice versa, in
qualitative agreement with observations (e.g.,
\citealt{Alexander08,Targett12}; also \citealt{Volonteri11}).

The rise of the specific growth of galaxies with redshift has recently
been attributed to the strong increase in the gas fractions of
galaxies from $z=0$ to 2 (\citealt{Daddi08, Daddi10, Tacconi10,
  Geach11}).  Given that the cosmological growth rate between SMBHs
and \mgal\ remains roughly constant, it seems that gas fractions also
play an important role in driving SMBH growth during this epoch.
However, clarifying the physical processes (feedback, volume effects,
etc.) that set \mdot/SFR $\approx10^{-3}$ and determine how gas
fraction translates to different sSFR and \smdot\ distributions
(Fig. \ref{GrowthSketch}a) remains an open issue that is beyond the
scope of this letter.

Our results provide insights into how the relationships between SMBHs
and their host galaxies are forged.  The vast majority (i.e.,
$\approx$98\%) of galaxies that form our parent sample are
main-sequence (MS) SFGs (e.g., \citealt{Rodighiero11}).  Morphological
and dynamical studies do find evidence of mergers among these galaxies
(e.g., \citealt{Elmegreen07, ForsterSchreiber09}), but being on the MS
(\citealt{Kartaltepe12}) implies their star-formation is not strongly
enhanced by these interactions (\citealt{DiMatteo08}).  Such MS
galaxies are responsible for $\approx$90\%\
(\citealt{Elbaz11,Rodighiero11,Sargent12}) of all star-formation
taking place during this epoch.  Of course, the most massive, distant
galaxies and SMBHs probed here will have grown their mass at earlier
times when other processes -- such as major-mergers -- may have played
a more dominant role. However, for SFGs to have
$\langle\frac{\dot{M}_{\rm BH}}{\rm SFR}\rangle\approx\frac{M_{\rm
    BH}}{M\ast}|_{z=0}$ during the time when the bulk of today's
stellar and SMBH mass was built-up implies that a significant fraction
of all SMBH growth takes place in MS galaxies whose SFRs are not
enhanced by mergers.  This view is consistent with recent studies of
the \mbox{sSFRs} and morphologies of X-ray selected AGN hosts which
find that the AGN population is dominated by non-mergers (e.g.,
\citealt{Cisternas11a, Schawinski11, Mullaney12, Kocevski12,
  Santini12}).  Indeed, results from recent hydrodynamical models
(e.g., \citealt{Bournaud11}) suggest that it is possible to have
efficient SMBH accretion inside gas rich, high redshift clumpy
galaxies, without invoking galaxy-galaxy interactions or mergers.

\vspace{3mm}
\noindent
We thank M. Dickinson, R. Gilli, A. Renzini and the anonymous referee.
We acknowledge financial support from STFC (DMA) and grants: ERC-StG
UPGAL 240039, ANR-08-JCJC-0008.


\end{document}